\documentclass[namedreferences]{kluwer}    
\newdisplay{guess}{Conjecture}
\usepackage{graphics}
\usepackage[mathcal]{eucal}
\usepackage{amssymb} 

\newcommand{\apj} {ApJ}
\newcommand{\mnras} {MNRAS}
\newcommand{\aap} {A\&A}

\newcommand{\deriv} [2] {\frac {d #1 } {d #2} }

\newcommand{\eqn} [1] {
\begin{equation}#1
\end{equation}}

\newlength{\lenA} %
\renewcommand{\title}[1]{\Large {\bf #1}}
\renewcommand{\author}[1]{\large #1}

\begin{document}                                                                                   
\begin{article}

\begin{opening}         
\title{Seismic diagnostics on stellar convection treatment
          from oscillation amplitudes of $p$-modes.} 
\author{R.~Samadi$^{(1,2)}$}  
\author{M.J.~Goupil$^{(2)}$}
\author{Y.~Lebreton$^{(2)}$}
\author{{\AA}.~Nordlund$^{(3)}$}  
\author{F.~Baudin$^{(4)}$}

\institute{(1) Astronomy Unit, Queen Mary, University of London, London, UK.
\\(2) LESIA, Observatoire de Paris-Meudon, Meudon, France.
\\(3) Niels Bohr Institute for Astronomy Physics and Geophysics, Denmark
\\(4) Institut d'Astrophysique Spatiale, Orsay, Fance.}

\runningauthor{R. Samadi}
\runningtitle{Seismic diagnostics on stellar convection treatment}
\date{}

\begin{abstract}
The excitation rate $P$ of solar $p$-modes 
 is computed with a  model of stochatic excitation which involves
constraints on the   averaged properties of the solar turbulence.These constraints 
 are obtained from a 3D simulation. 
Resulting values for $P$ are found $\sim 9$ times larger 
than when the calculation  assumes   properties of turbulent convection which are derived 
 from an 1D solar model based on \inlinecite{Gough77}'s formulation of the  mixing-length theory (GMLT).
This difference is mainly due to the assumed values for the mean anisotropy of the velocity field in each case.
Calculations based on 3D constraints  bring the $P$ maximum  closer to the
 observational one.
\\We also compute  $P$ for several models of intermediate mass stars 
($1 \lesssim M \lesssim 2~M_\odot$).
Differences  in the values of $P_{\rm max}$ between models computed with 
the classical  mixing-length theory and GMLT models are found  large enough
 for main sequence stars 
 to suggest that measurements 
of $P$ in this mass range will be able 
to discriminate between different models of turbulent   convection. 
\end{abstract}
\keywords{convection, turbulence, oscillations, Sun}

\end{opening}           

\vspace{-0.6cm}

\section{Introduction}

Excitation of solar-type oscillations is attributed to turbulent 
movements in the outer convective zone of intermediate mass stars.

Accurate measurement of the  rate $P$ at which 
acoustic energy is  injected into such oscillations will 
be possible with the future seismic missions  (e.g. COROT and EDDINGTON).
Comparison between measured and theoretical values of $P$ obtained 
with different models of turbulent convection will then 
provide valuable information about the properties of stellar convection zones.

Models for  stochastic excitation have been proposed 
by several authors (e.g. \opencite{GK77}; \opencite{Balmforth92c};  \opencite{Samadi00I}).
In  the present work, we consider the formulation of \citeauthor{Samadi00I} (2001, Paper~I hereafter).
Constraints on the time averaged properties of the solar turbulent medium are obtained 
from a 3D simulation. They allow  us to compute $P$ and compare it with 
solar seismic observations and results obtained with \inlinecite{Gough77}'s
 formulation of the mixing-length theory (GMLT hereafter).

\section{The model of stochastic excitation}

According to Paper~I, the rate at which a given mode with frequency $\omega_0$ is excited can be written  as:
\eqn{
\hspace{-0.18cm} P(\omega_0)   \propto  
\int_{0}^{M}{\rm d}m \,  {\Phi \over 3}  \, \rho_0 w ^4 \, \left \{ {16 \over 15} \, \frac{\Phi}{3} \, \left (\deriv { \xi_r} {r} \right )^2 \,  S_R 
+   { 4 \over 3} \,
\left ( \frac{\alpha_s \, \tilde s}{\rho_0 \, w} \right )^2  \,   \frac{g_{\rm r}  }{\omega_0^2} \,   S_S \right \}
\label{eqn:P}
} %
In Eq.~(\ref{eqn:P}), $\rho_0$ is the mean density, $\displaystyle{\xi_{\rm r}}$ is the
 radial component of the fluid displacement adiabatic eigenfunction ${\bf \xi}$,   $\displaystyle{\alpha_s =\left ( \partial p /\partial s  \right )_\rho}$ 
where $p$ denotes the  pressure and $s$ the entropy,
 $\tilde s^2$ is the rms value of the entropy fluctuations, $g_r(\xi_{\rm r},m)$ involves the first and the second derivatives of  $\bf \xi$ with respect to $r$,  $S_R(\omega_0,m)$ and $S_S(\omega_0,m)$ are driving sources inferred from the Reynolds and the entropy fluctuations respectively,
$\Phi$ is \inlinecite{Gough77}'s  mean anisotropy factor defined as $ \Phi(m) \equiv 
{<{\bf u}^2> - <{\bf u}>^2} / {w^2}$ 
where $\bf u$ is the velocity field, $< . >$ denotes time and horizontal  average and $w$  is the mean vertical velocity ($w^2 \equiv <u_z^2>-<u_z>^2$). Expressions for  $S_R(\omega_0,m)$ , $S_S(\omega_0,m)$ and $g_r(\xi_{\rm r},m)$  are given in Paper~I.

The driving sources $S_R$ and $S_S$  include the   turbulent kinetic energy spectrum $E(k,m)$, the turbulent spectrum of the entropy fluctuations $E_s(k,m)$ and  $\chi_k(\omega)$ the frequency-dependent part of the turbulent spectra which model the correlation time-scale of an eddy with wavenumber $k$. 
$\chi_k(\omega)$ is modeled here with a non-gaussian function constraint from the 3D simulation.

\section{Numerical constraints and computation in the solar case}

We consider a 3D simulation of the upper part of the solar convective zone as obtained by  \inlinecite{Stein98}. The simulated domain is 3.2 Mm deep and its surface is 6 x 6 ${\rm Mm}^2$. 
The grid of mesh points is 256 x 256 x 163, the total duration 27 mn and the sampling time 30s.

The simulation data are used to determine the quantities $E(k,m)$,  $E_s(k,m)$, $w$, $\tilde s$ 
and $\Phi$ involved in the theoretical expressions for  $S_R$, $S_S$ and $P$.
More details will be given in  forthcoming papers. 



We compute $P$ according to Eq.~(\ref{eqn:P}):
the eigenfunctions ($\xi$) and their frequencies ($\nu=\omega_0/2\pi$) are 
calculated with \inlinecite{Balmforth92c}'s non-adiabatic code for a solar 1D model 
built with the  GMLT approach.
The $k$-dependency of $E(k,z)$,  is modeled as $(k/k_0)^{+1}$ for $k < k_0$ and 
as $(k/k_0)^{-5/3}$ for $k>k_0$ where 
$k_0 = 2 \pi /  \beta \Lambda$, $\Lambda=\alpha H_p$ 
is the mixing-length, $H_p$ the pressure scale height.
The value of  the mixing-length parameter, $\alpha$, is imposed by 
a solar calibration  of the 1D GMLT model. The value of $k_0$ -~hence of $\beta$~-
 is obtained from the 3D simulation.  The above analytical $k$-dependency of $E$ and   $E_s(k,z)$ 
reproduce the global features of $E$ and $E_s$ derived  from the  3D simulation.

\begin{figure}[h]
\begin{center}
\vspace{-0.5cm}
\resizebox{10.5cm}{!}{\includegraphics{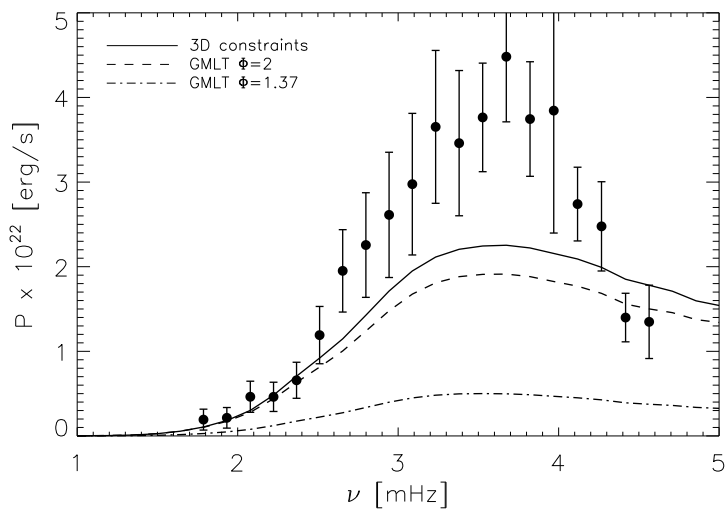}}
\end{center}
\vspace{-0.7cm}
\caption{Solar $p$-modes excitation rate $P(\nu)$ (See text for details).}
\label{fig:cmp_pow_gh}
\end{figure}

Results of $P$ computations are presented 
in Fig.~\ref{fig:cmp_pow_gh} ({\bf solid curve}) and 
compared with  values of $P$ ({\bf filled dots}) 
derived from solar seismic measurement by \inlinecite{Chaplin98}.



We also compute  $P$ in the case when the quantities involved in Eq.~(\ref{eqn:P}) are all 
obtained from the GMLT model. In  the mixing-length approach, the anisotropy factor, $\Phi$, is a free parameter.
We then assume two different values for $\Phi$: $\Phi=1.3745$ ({\bf dot dashed curves}) and $\Phi=2$  ({\bf dashed curves}).
The value $\Phi=1.3745$ used in the GMLT model provides the best fit  between computed solar 
damping rates and the measured ones by \inlinecite{Chaplin98}. 

\vspace{-0.3cm}

Without any  adjustement of scaling parameters 
and instead using all the constraints inferred from the 3D simulation, 
we find a maximum of $P$ much larger ($\sim 9$ times larger) than is obtained 
using a 1D GMLT solar model  with $\Phi=1.37$. 
This difference is mainly due to the assumed value for $\Phi$  in the excitation region.
Indeed, analysis of the 3D simulation suggests that $\Phi \sim 2$ in the excitation region.
When  $\Phi=2$ is used to compute the excitation rate with the GMLT model, 
$P$ comes close to $P$ calculated with the 3D simulation constraints.

Our result then shows  that the  values of 
$\Phi$ found for the solar GMLT model when adjusted on the damping rates
 is  not compatible 
with the actual properties of turbulent medium 
in the excitation region. 
An improvement could come from a consistent calculation 
which would  assume a depth dependent $\Phi(m)$, 
as suggested by the simulation, in both  damping  and
excitation rates computation. 

Our calculations using the 3D constraints  bring $P_{\rm max}$, the $P$  maximum,
 closer to the solar  seismic  measurements but still  under-estimates them by a factor $\sim 2$.
More sophisticated assumptions for $k_0(z)$ and  $\lambda$ will likely lead to a better  agreement with the observations.


\section{Scanning the HR diagram}

We consider two sets of stellar models previously investigated  by \inlinecite{Samadi01b}:
Models in the first set are computed with the classical MLT whereas those belonging to the
 the second set are computed with GMLT. We find that  the maximum of $P$,  
 $P_{\rm max}$,  scales as $3.2 \, {\rm log}(L/L_\odot \times M_\odot / M)$ 
for the first set and scales as  $3.5 \, {\rm log}(L/L_\odot \times M_\odot / M)$
 for the second one ($L$, $L_\odot$, $M$ and $M_\odot$ have their usual meaning).
This result suggests that measurements of $P$ in several 
differents intermediate mass stars ($1 \lesssim M \lesssim 2~M_\odot$) 
will enable one to discriminate between different models of turbulent   convection.

\vspace{-0.3cm}

\acknowledgements

\vspace{-0.2cm}

RS acknowledges support by the Particle Physics and Astronomy Research Council of the UK under the grant PPA/G/O/1998/00576.
We thank Guenter Houdek for providing us the solar model.

\vspace{-0.5cm}

\bibliographystyle{klunamed}

\end{article}
\end{document}